\documentclass[aps,prl,superscriptaddress,amsmath,amssymb,floatfix,twocolumn]{revtex4}

\usepackage{graphicx}
\usepackage{epstopdf}
\usepackage{dcolumn}
\usepackage{bm}
\usepackage{color}
\usepackage{ulem}
\usepackage{epsfig,float,afterpage,amssymb,wrapfig,psfrag}
\usepackage[caption=false]{subfig}
\usepackage{placeins}
\usepackage{comment}
\DeclareMathAlphabet{\bi}{OML}{cmm}{b}{it}

\newcommand{\be}[0]{\begin{equation}}
\newcommand{\ee}[0]{\end{equation}}
\newcommand{\ba}[0]{\begin{eqnarray}}
\newcommand{\ea}[0]{\end{eqnarray}}

\begin{document}
\title{Dynamical scaling of charge and spin responses at  a Kondo destruction quantum critical point}

\author{Ang Cai}
\affiliation{Department of Physics and Astronomy, Rice Center for Quantum Materials,  Rice University,
Houston, Texas, 77005, USA}
\author{Zuodong Yu}
\address{Zhejiang Institute of Modern Physics, Zhejiang University, Hangzhou,  Zhejiang 310058, China}
\address{Zhejiang Province Key Laboratory of Quantum Technology and Devices, Zhejiang University, Hangzhou 310027, China}
\author{Haoyu Hu}
\affiliation{Department of Physics and Astronomy, Rice Center for Quantum Materials,  Rice University,
Houston, Texas, 77005, USA}
\author{Stefan Kirchner}
\email{kirchner@correlated-matter.com}
\address{Zhejiang Institute of Modern Physics, Zhejiang University, Hangzhou,  Zhejiang 310058, China}
\address{Zhejiang Province Key Laboratory of Quantum Technology and Devices, Zhejiang University, Hangzhou 310027, China}
\author{Qimiao Si}
\email{qmsi@rice.edu}
\affiliation{Department of Physics and Astronomy, Rice Center for Quantum Materials,  Rice University,
Houston, Texas, 77005, USA}

\date{\today}
\begin{abstract}
Quantum critical points
often arise in metals perched at the border of an antiferromagnetic order.
The recent observation of 
singular and dynamically scaling charge conductivity
in an antiferromagnetic quantum critical heavy fermion metal  implicates beyond-Landau quantum criticality.
Here we study the charge and spin dynamics of a Kondo destruction quantum critical point (QCP), 
as realized in an SU(2)-symmetric
 Bose-Fermi Kondo model.
We find that the critical exponents and scaling functions 
 of the spin and single-particle responses of the QCP in the SU(2) case are essentially the same as those 
 of the large-N limit, 
showing that $1/N$ corrections are subleading. Building on this insight, 
 we demonstrate that the charge responses
  at the Kondo destruction QCP are singular and 
 obey $\omega/T$ scaling.
 This property persists at the Kondo destruction QCP of the SU(2)-symmetric Kondo lattice 
 model.
 \end{abstract}

\date{\today}

\maketitle

{\it Introduction.~}
Quantum criticality is of extensive current interest to a variety of strongly correlated systems
\cite{QCNP2013,Si_Science10,Coleman-Nature,Sachdev-book}.
Within the Landau framework, phases of matter are differentiated by the spontaneous breaking 
of global symmetry and its associated order parameter, and quantum criticality is described 
by the fluctuations of the order parameter. For a continuous transition between
antiferromagnetic to paramagnetic phases at $T=0$, the corresponding singularity is 
associated with the slow fluctuations 
of the staggered magnetization \cite{hertz1976quantum}. 

Antiferromagnetic (AF) heavy fermion metals provide a prototype setting to                                                                  
elucidate the quantum critical properties and the associated strange-metal physics. In these systems,
strong correlations manifest themselves through the development of
local moments out of their $f$-electrons.
The local moments interplay with a band of conduction electrons by an AF Kondo coupling, 
and interact with each other via an RKKY coupling.
In the process of understanding heavy fermion quantum criticality, 
it has been emphasized that the Landau framework, in the form of
a spin-density wave (SDW)
QCP \cite{hertz1976quantum, millis1993effect, moriya2012spin}, can break down in a fundamental way.
The beyond-Landau physics has been characterized in terms of
the notion of Kondo destruction  
\cite{Si-Nature,Colemanetal,senthil2004a}.
The distinction
of the Kondo destruction quantum criticality from its SDW counterpart
 reflects
the amplitude of the Kondo singlet going to zero as the system approaches the AF QCP 
from the paramagnetic phase. 
 Correspondingly, 
 the quasiparticle weight 
 vanishes at the QCP and
the
 Fermi surface jumps across the transition.
 
In the context of critical phenomenon,  the Kondo destruction QCP epitomizes
the effect of quantum entanglement on criticality singularity. 
From the perspective of strongly correlated electrons, it 
corresponds to a partial Mott transition, {\it i.e.}
the localization of the $4f$-electrons.
Such an
electronic localization-delocalization transition 
links quantum critical heavy fermion metals 
to other strongly correlated systems. For instance, in the cuprate superconductors
near optimal  hole-doping, Hall measurements implicate an electron 
localization-delocalization transition \cite{Badoux16}.  In an organic superconductor,
such measurements have suggested a similarly rapid change in the carrier density \cite{Oike15}.
Finally, in the twisted bilayer graphene, quantum oscillation measurements indicate a small Fermi surface 
when the system
is doped away from the half-filled correlated insulator \cite{Cao18}.

In quantum critical heavy fermion metals, there
is 
extensive experimental evidence 
for
the Fermi surface jump
\cite{paschen2004,Friedemann.10,shishido2005}
as well as the emerging Kondo destruction energy scale \cite{paschen2004,Gegenwart2007}.
One of the early experimental clues for anomalous heavy fermion quantum criticality
came from the observation of  $\omega/T$ scaling together with an anomalous value for the critical exponent
 in the spin dynamics \cite{schroder}. 
The Kondo destruction quantum criticality has provided a natural 
 understanding of such singular dynamic scaling in the critical spin response \cite{Si-Nature,si2014kondo}. 

Recently, terahertz spectroscopy measurements in a quantum critical heavy fermion metal 
have discovered 
a charge response that is singular and satisfies $\omega/T$ scaling \cite{prochaska2018singular}.
This is inconsistent with an SDW QCP, where only the response of antiferromagnetic order parameter 
should be singular and the charge correlations are expected to be smooth.
A critical destruction of the Kondo effect, however,
involves the localization-delocalization of the $f$-electrons at the QCP. 
Thus, the charge degrees of freedom are an integral part of the quantum criticality 
leading to the suggestion of a singular response
 in the charge channel.
Indications of a singular charge response have appeared 
in a dynamical large-$N$ study (see below for the definition of $N$)
for a Kondo destruction QCP of the Bose-Fermi Kondo model 
(BFKM) and in related settings \cite{zhu2004quantum,Kirchner.05a,Pix12.1,Komijani2019},
with 
the BFKM being associated with the Kondo lattice model within the approach of
extended dynamical mean field theory (EDMFT).
In light of the recent experimental development, 
theoretical studies at the physical $N=2$ case
are called for.

In this letter, we demonstrate  for the first time
that  the charge response of
the Kondo destruction QCP is singular and has dynamical $\omega/T$ scaling 
in the physical $N=2$ case.
Our result is facilitated by analyzing the BFKM at both $N=2$ and a dynamical large-$N$ limit,
which shows
that $1/N$ corrections to the scaling quantities are subleading but not dangerously irrelevant. 
Based on this insight, 
we carry out calculations on both
the BFKM and the Kondo lattice model.
Our results provide the theoretical basis to understand
the striking recent measurement of singular charge response
at an antiferromagnetic heavy fermion QCP \cite{prochaska2018singular}.

{\it BFKM with SU(2) symmetry.~}
We will study the quantum critical properties of the spin rotationally invariant BFKM
\cite{zhu2002,zarand2002,smith1999,sengupta2000} and
related Bose-Fermi 
Anderson model (BFAM),
compare the results determined for the SU(2)-invariant case ($N=2$)
with those obtained
in the dynamical large-N limit.
For the SU(2)  case, we will study the BFAM defined by the following Hamiltonian:
\begin{eqnarray}
{\cal H}_{\text{BFA}}
&=& \sum_{\sigma} \epsilon_{d} d_{\sigma}^{\dagger}d_{\sigma} + U n_{\uparrow} n_{\downarrow}  \nonumber  \\
&+& \sum_{p\sigma} \left( V d_{\sigma}^{\dagger} c_{p\sigma} + h.c. \right) + 
\sum_{p\sigma} \epsilon_{p}~c_{p\sigma}^{\dagger}~ c_{p\sigma}
 \nonumber \\
&+& \; g \sum_{p} {\bf S}_{d} \cdot
{\bf \Phi}
+ 
 \sum_{p} w_{p}\,{\bf \Phi}_{p}^{\;\dagger}\cdot {\bf \Phi}_{p} \, .
\label{H-BFA-su2}
\end{eqnarray}
Here, strongly correlated $d$-electrons, with the Hubbard interaction $U$ defined in terms of 
$n_{\sigma}=d^{\dagger}_{\sigma} d_{\sigma}$ and in the presence of a particle-hole symmetry, 
$\epsilon_d = -U/2$,
hybridize with the conduction $c$-electrons
with an amplitude $V$. For the interactions we consider, the hybridization amounts 
to a Kondo-coupling of
the $d$-electron spin,
${\bf S}_{d}
=d^{\dagger}_{\sigma} \bm{\tau}_{\sigma \sigma^{\prime} } d_{ \sigma^{\prime} }$, 
with $\bm{\tau}_{\sigma \sigma^{\prime} }$ being the three component Pauli matrices,
to the fermionic $c$-bath. Simultaneously, the $d$-electron spin is coupled to a 
vector ${\bf \Phi}$-bosonic bath; we have defined
${\bf \Phi} = \sum_p ( {\bf \Phi}_{p} +  {\bf \Phi}_{-p}^{\;\dagger} )$.
We assume a flat fermionic density of states
\begin{equation}
\rho_{f}(\epsilon)=\sum_{p} \delta(\epsilon-\epsilon_{p}) = \rho_{0} \Theta (D-\epsilon) 
\Theta (D+\epsilon)
\, ,
\label{eq:rho-f}
\end{equation}
where $\Theta$ is the Heaviside function. This 
defines a hybridization function $\Gamma(\epsilon)=\Gamma=\pi\rho_{0}V^{2}$ for $\epsilon \in (-D,D)$.
We choose $D=1$ as the energy unit.
 For the bosonic bath, we consider a subohmic spectrum ($s<1$)
 \begin{equation}
\rho_{b}(\omega)=\sum_{p} \delta(\omega-\omega_{p} ) = K_{0} \omega^{s} e^{-\omega/\Lambda} \Theta(\omega) \, .
\label{eq:rho-b}
\end{equation} 
where  $\Lambda$ is a cutoff frequency.
The model 
is
studied using a 
continuous-time Quantum Monte Carlo (CT-QMC) method developed in 
Ref.\,\cite{cai2019} (see also Refs.\,\cite{Otsuki-2013,pixley2011,pixley2013,werner2006prl,werner2006prb}).

{\it BFKM in dynamical large-$N$ limit.~}
The BFKM in the dynamical large-$N$ limit is defined in terms of the Hamiltonian:
\begin{eqnarray} 
{\cal H}_{\text{BFK}} &=& 
({J }/{N}) 
\sum_{\alpha}{\bf S} 
\cdot {\bf s}_{\alpha} 
+ \sum_{p,\alpha,\sigma} \epsilon_{p}~c_{p \alpha 
\sigma}^{\dagger} c_{p \alpha \sigma} 
\nonumber\\ 
&+& 
({g}/{\sqrt{N}})
{\bf S} \cdot 
{\bf \Phi}
+ \sum_{p} 
w_{p}\,{\bf \Phi}_{p}^{\;\dagger}\cdot {\bf \Phi}_{p}. 
\label{H-MBFK} 
\end{eqnarray} 
As in the SU(2) case, a local moment ${\bf S}$ is coupled to
a  fermionic and a vector bosonic bath,
$c_{p\alpha\sigma}$ and
${\bf \Phi}_p$ respectively. The spin symmetry is SU(N),
with $\sigma = 1, \ldots, N$,
and the channel symmetry is SU($\kappa N$), with
$\alpha=1, \ldots, \kappa N$
(Ref.\,\cite{Cox.93,Parcollet.98}). 
Here, ${\bf \Phi}$
has $N^2-1$
components. 
The density of states is likewise given by
Eqs.~(\ref{eq:rho-f},\ref{eq:rho-b}).
The bare bath Green's functions are 
${\cal G}_0 = - \langle
T_{\tau} c_{\sigma\alpha}(\tau) c_{\sigma\alpha}^{\dagger}(0)
\rangle _0$ and 
${\cal G}_{\Phi} =  \langle
T_{\tau} \Phi(\tau) \Phi^{\dagger}(0)
\rangle _0$.

We use a fermionic spinon representation,
 $S_{\sigma 
\sigma'}=f_{\sigma}^{\dag}f_{\sigma'}-
\delta_{\sigma, \sigma'}/2$, 
enforcing  the constraint of the Hilbert space
$\sum_{\sigma=1}^N 
f_{\sigma}^{\dag}f_{\sigma}= 
N/2$ by a Lagrange multiplier $i\mu$. 
The conduction electrons are 
in the fundamental representation of the SU($N$)$\times$SU($\kappa N$)
group.
A
 dynamical 
field $B_{\alpha}(\tau)$ is used to decouple 
the Kondo coupling,
$(J/N)\sum_{\sigma \sigma'} \left( 
f_{\sigma}^{\dag}f_{\sigma'}- 
\delta_{\sigma, \sigma'} /2
\right) c_{\alpha\sigma'}^{\dagger}c_{\alpha\sigma} $,
leading to 
a $B_{\alpha}^{\dag}\sum_{\sigma}c_{\alpha 
\sigma}^{\dag}f_{\sigma}/\sqrt{N}$ interaction.
The $B$-field is charge-carrying, given that the spinon field $f$ is charge-neutral.

Taking the large-$N$ limit with $\kappa$ being kept fixed leads to the following saddle-point equations:
\begin{eqnarray} 
G_B^{-1}( i\omega_n) &=& 
1/J
- \Sigma_B( i\omega_n);
~~~
\Sigma_B(\tau) = - {\cal G}_{0}(\tau) G_f(-\tau)
 \nonumber \\
G_f^{-1}(i\omega_n)& =& i\omega_n - \lambda -
\Sigma_f(i\omega_n) ; \nonumber \\ 
\Sigma_f(\tau)&=& \kappa {\cal G}_{0}(\tau) G_B(\tau) + g^2 
G_f(\tau){\cal G}_{\Phi}(\tau) ,
\label{NCA}
\end{eqnarray} 
which  are supplemented by the following constraint:
\begin{eqnarray} 
G_f(\tau = 0^{-}) = ({1}/{\beta}) \sum_{i\omega_n}
G_f (i\omega_n) {\rm e}^{i\omega_n 0^+} = {1}/{2} .
\label{constraint}
\end{eqnarray} 
These equations are solved on the real frequency axis.
For definiteness, we will fix $\kappa=1/2$.

{\it Critical properties -- dynamical large-$N$ limit {\it vs.} SU(2).~}
In the large-$N$ limit, a QCP separates the strong-coupling Kondo phase from
a Kondo destruction critical phase. In the SU(2) model, for the value 
of $s$ we focus on, the Kondo destruction phase also corresponds to a 
critical phase \cite{cai2019}.
Comparing 
the critical properties of the dynamical large-$N$ limit 
 with the SU(2) model
allows us to assess the degree to which $1/N$ corrections modify
the leading quantum critical singularities.

\begin{figure}[t!]
\captionsetup[subfigure]{labelformat=empty}
  \centering
    \mbox{\includegraphics[width=0.98\columnwidth]{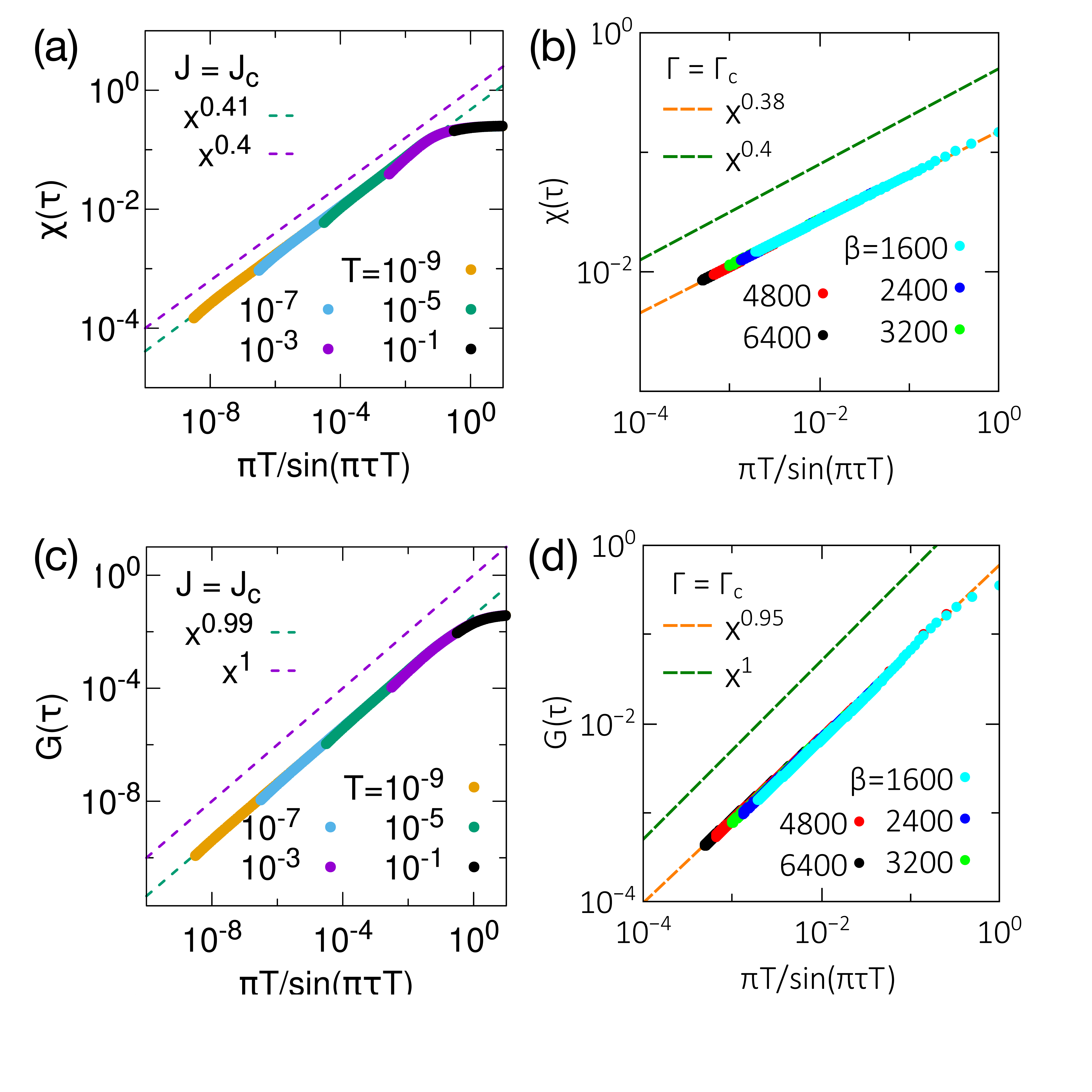}}    
\caption{Local spin susceptibility $\chi(\tau)$ at the QCP of (a) the dynamical large-$N$ limit (purple dashed line represents the analytically obtained leading $T=0$ behavior)
 and (b)  the SU(2) case; 
 and 
 electron Green's function $G(\tau)$ at the QCP of (c) the dynamical large-$N$ limit and (d) 
the SU(2) case.
 In (a,c), the temperature $T$ is measured
 in $D$.
In (b,d), $\beta=1/T$.}
\label{fig:chi}
\end{figure}

We first consider the local spin susceptibility, $\chi$, at the QCP and in the Kondo destruction
 phase. In the dynamical large-$N$ limit,
\begin{eqnarray}
\chi(\tau) = -G_f(\tau) G_f(-\tau) \, .
\label{eq:chi-from-Gf}
\end{eqnarray}
In the SU(2) case, $\chi(\tau)$ is directly calculated from the CT-QMC procedure.
The result for the dynamical large-$N$ calculations for $s=0.6$ ({\it i.e.}, $\epsilon=1-s=0.4$)
is shown in 
Fig.~\ref{fig:chi}(a). We find that $\chi$ as a function of the imaginary time, $\tau$,  collapses
in terms of $\pi T/\sin(\pi \tau T)$, where $T$ is the temperature, with a power-law exponent $\eta$
that is 
less than $1$. This implies a singular spin response: The static local spin susceptibility
diverges in the $T \rightarrow 0$ limit, and so does the $T=0$ local spin susceptibility as $\omega \rightarrow 0$;
both divergencies have the power-law exponent of $1-\eta$.
The exponent $\eta$ is numerically fit to be $0.41$. This value 
is in excellent agreement with the analytical result,
$\eta=\epsilon=0.4$, that can be 
extracted from the saddle point equations (\ref{NCA},\,\ref{constraint}) in the zero-temperature limit \cite{zhu2004quantum}. 

As a comparison, we show 
in Fig.\,\ref{fig:chi}(b)
the CT-QMC result for  $\chi(\tau)$ at the QCP of the SU(2) BFAM,
again for $s=0.6$.  
Unlike the large-$N$ limit where the real-frequency analysis is carried out over many (more than 10) 
decades, 
here the dynamical range is more limited.
Still, by using the
algorithm recently developed in Ref.\,\cite{cai2019},
we are able to reach low-enough temperatures and a sufficiently large dynamical range in $\tau$
to determine the scaling properties in the quantum critical regime. We see from 
Fig.\,\ref{fig:chi}(b) that the scaling function is also a power-law of $\pi T/\sin(\pi \tau T)$.
The fitted exponent is $0.38$, which
is quite close to the large-$N$ result ($0.41$ as calculated and $0.4$ as expected). 
We attribute the difference to the subleading corrections
that 
are amplified in the CT-QMC calculation, given
the
narrower scaling range 
being accessed.

\begin{figure}[t!]
\captionsetup[subfigure]{labelformat=empty}
  \centering
      \mbox{\includegraphics[width=0.98\columnwidth]{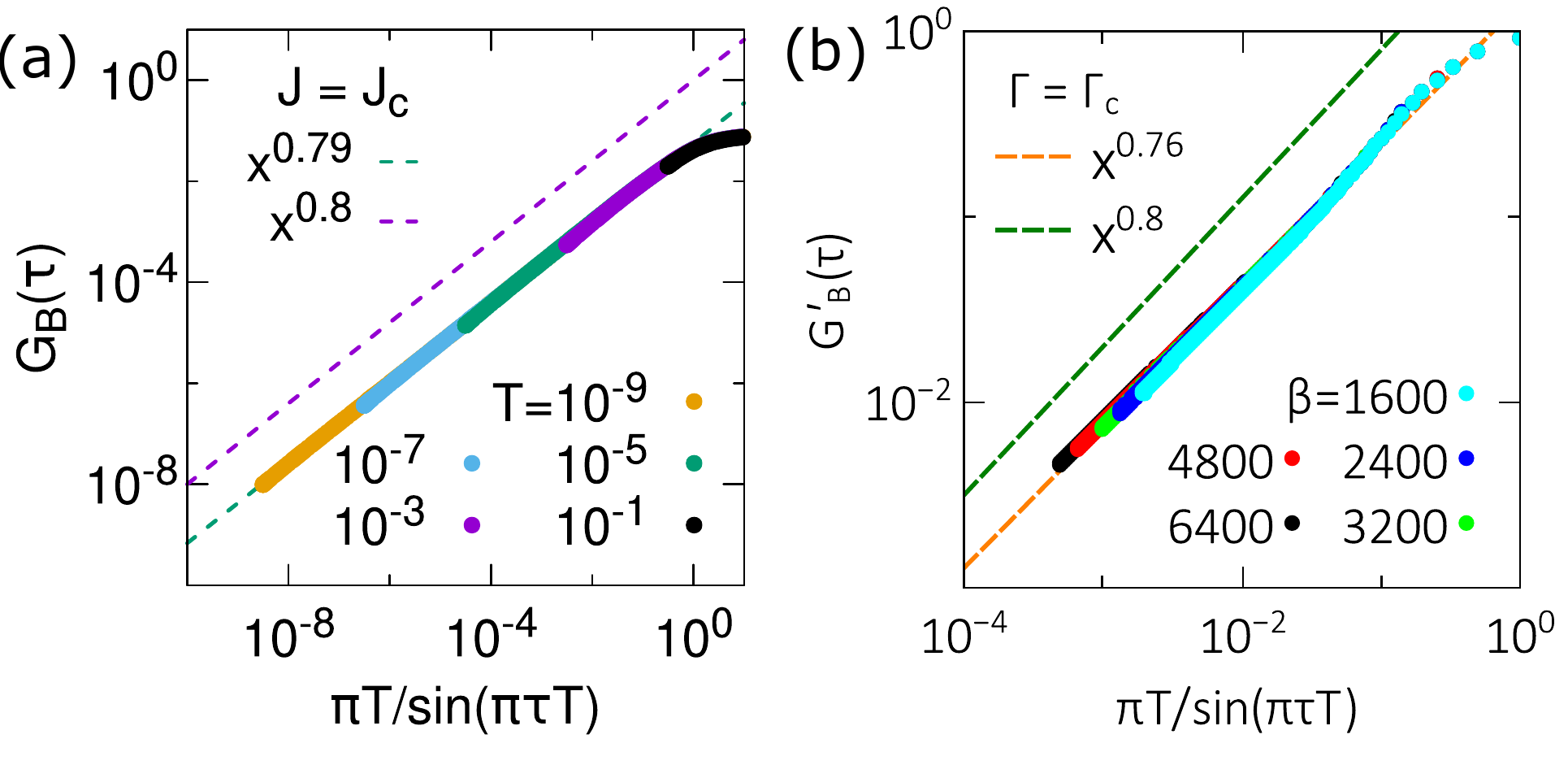}}    
\caption{
Singular charge response of the $B$-field calculated in the dynamical large-$N$ limit (a) and 
extracted in the SU(2) case (b) obtained at the QCP.
In (a), the temperature $T$ is in unit of 
$D$.
In (b), $\beta=1/T$.
}
\label{fig:GB_qcp}
\end{figure}

We now turn to a 
parallel study of  the $d$-electron Green's function $G(\tau)$. In the
large-$N$ limit, it is determined as follows:
\begin{eqnarray}
G(\tau) = G_f(\tau) G_B(-\tau) \, .
\label{eq:G-from-GfGB}
\end{eqnarray}
For the SU(2) case, $G(\tau)$ is again directly calculated from the CT-QMC procedure. The results 
for the QCP is shown for the large-$N$ limit in Fig.\,\ref{fig:chi}(c), with the exponent being $0.99$,
very close to the value analytically expected, which is $1$. In addition, for the SU(2) case shown in 
Fig.\,\ref{fig:chi}(d),
within
numerical accuracy, both the critical exponent and the scaling functions
are essentially the same as the large-$N$ limit.

{\it Critical charge and spin responses and $\omega/T$ scaling.~}
The above calculations and comparisons lead to an important new insight. 
For the Kondo destruction QCP, the leading critical singularities determined
in the dynamical large-$N$ limit applies to finite $N$ including $N=2$.
This implies that, while the
$1/N$-corrections modify the location of the quantum 
critical point, they preserve its
Kondo-destruction nature and, equally important, make only subleading contributions to the critical singularities.
Analyzing the Feynman diagrams shows that the processes at the $1/N$ and higher orders are irrelevant \cite{Cox.93}. 
In addition, the $1/N$ corrections cannot be dangerously irrelevant:
Given that the susceptibilities at the large-$N$ limit satisfy $\omega/T$ scaling (see below), 
the subleading corrections will preserve the 
leading singularities as a function of not only the frequency but also the temperature.
This insight leads to
a remarkable simplification, because it implies that we can
use the results determined in the dynamical large-$N$ limit to gain an understanding about the critical properties at 
realistic  Kondo destruction QCPs at finite $N$.

We start from the response of the charge-carrying $B$-field, which is expected 
to be singular~\cite{zhu2004quantum}. 
In Fig.\,\ref{fig:GB_qcp}(a), we show that it too is a power law of 
 $\pi T/\sin(\pi \tau T)$, with a critical exponent being
  very close to the value determined analytically for the leading singularity,
{\it i.e.} $0.8$ (which corresponds to $1-\epsilon/2$).

The lack of $1/N$-corrections to the leading critical
 singularities at the Kondo destruction QCP suggests that the structure 
of Eqs.~(\ref{eq:chi-from-Gf},\ref{eq:G-from-GfGB}) is still valid at finite $N$. The
 form of the scaling functions 
 simplifies these equations into 
$\chi(\tau) = [G_f(\tau) ]^2$ and $G(\tau) = G_f(\tau) G_B(\tau)$ [for $\tau \in (0,\beta)$].
We therefore
define
\begin{eqnarray}
G_B^{\prime}(\tau) = \frac{G(\tau)}{\sqrt{\chi(\tau)}}
\label{eq:GB'}
\end{eqnarray}
as a measure of the singular correlator of the charge-carrying $B$-field. 
The $\tau$-dependence of $G_B^{\prime}$ from our CT-QMC calculation of the SU(2) BFAM is presented in
Fig.\,\ref{fig:GB_qcp}(b). Both the critical exponent and the scaling function are,
within the numerical uncertainty,
the same as for the large-$N$ result.
This particular form of scaling function in the $\tau$-dependence, with its power-law exponent being less than $1$,
implies a singular dependence on $\omega$ and $T$ with an $\omega/T$ scaling.

\begin{figure}[t!]
\captionsetup[subfigure]{labelformat=empty}
  \centering
    \mbox{\includegraphics[width=0.48\columnwidth]{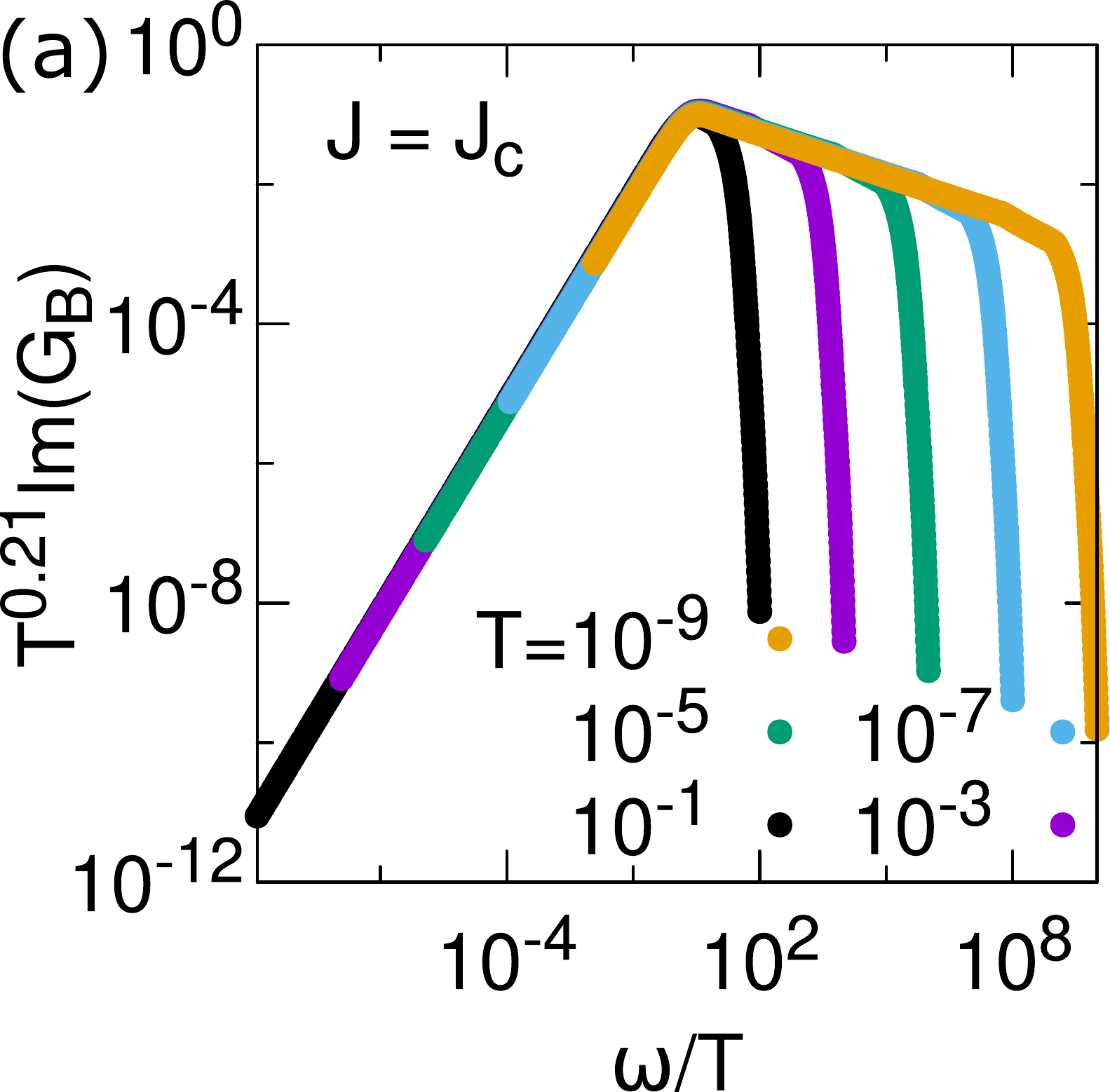}}    
    \mbox{\includegraphics[width=0.48\columnwidth]{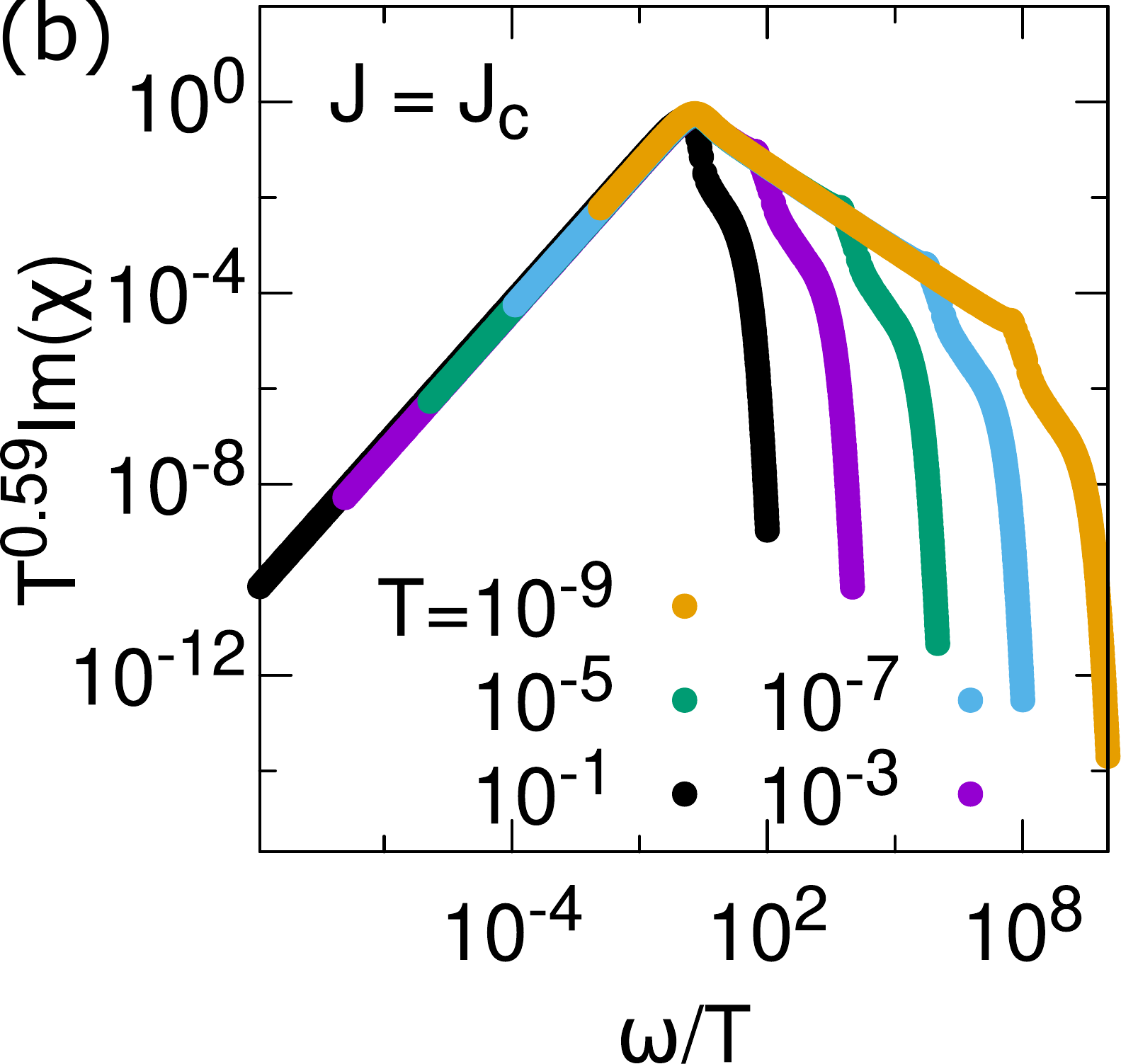}}
\caption{$\omega/T$ scaling at the QCP. (a) The charge response, 
showing the spectral function of the $B$-field and (b) the spin response, showing the spectral function of ${\bf S}$, 
both in the dynamical large-$N$ limit.
The temperature $T$ and $\omega$ are in unit of 
$D$.
}
\label{fig:GB_qcp_omegaT}
\end{figure}

Thus, we have established that the Kondo destruction QCP displays a singular response 
in both charge and 
spin channels. The real-frequency dependences of the spectral functions of 
both the charge-carrying $B$-field 
and 
the spin ${\bf S}$ are shown to collapse in $\omega/T$
in Figs.\,\ref{fig:GB_qcp_omegaT}(a,b). 
Each quantity satisfies $\omega/T$ scaling over 
a dynamical
range of more than 15 decades.

{\it Kondo lattice model.~} 
Since the lattice model is more relevant to the real materials, we study the 
SU(2)-symmetric Kondo lattice model.
The model itself is standard, as is the EDMFT approach \cite{si2014kondo}.
However, systematic calculations for the SU(2)-symmetric case has only
become possible recently with the advent of the SU(2) CT-QMC method \cite{cai2019}.
Within EDMFT the lattice model is described by 
the BFKM involving
self-consistently determined bath.

\begin{figure}[t!]
\captionsetup[subfigure]{labelformat=empty}
  \centering
    \mbox{\includegraphics[width=0.8\columnwidth]{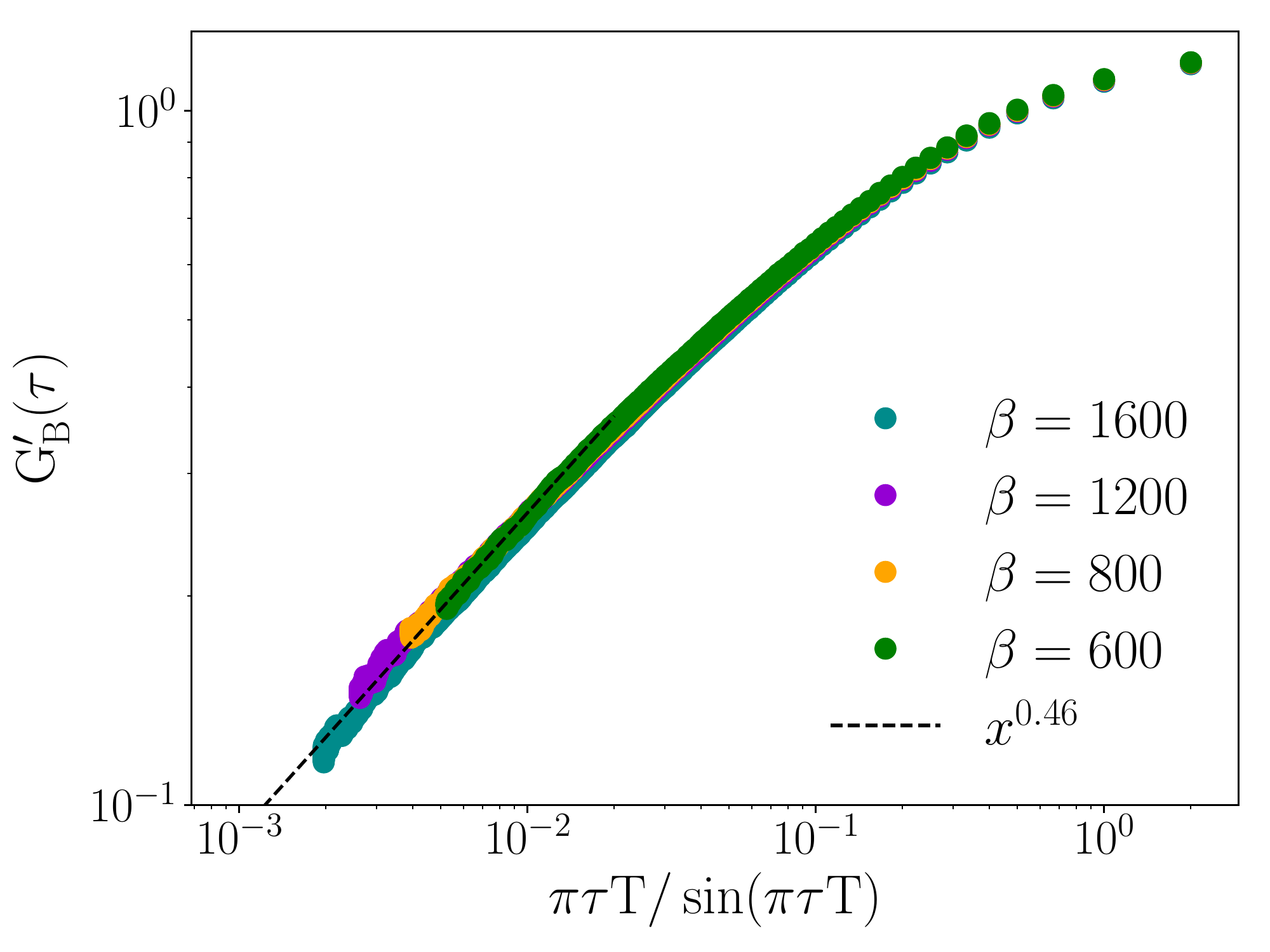}}    
\caption{
Singular charge response of the $B$-field calculated in the lattice model at the Kondo destruction QCP. 
}
\label{fig:self_consistent}
\end{figure}

In the lattice model, we numerically identity a Kondo destruction QCP,
which
 separates a paramagnetic Kondo screened 
 phase from an antiferromagnetic Kondo destruction phase \cite{SU2EDMFT}.
We then
 investigate the charge response $G_B'$ at the QCP. 
As shown in Fig.~\ref{fig:self_consistent}, we find
 $G_B'$ to collapse
 as a function of $\pi T/\sin(\pi \tau T) $.
The critical exponent is about 0.5 which corresponds to the $s\sim 0$ case in the BFKM.
This form of charge response is
 critical and satisfies $\omega/T$ scaling.\\

{\it Discussion and conclusion.~}
Importantly, in both the Bose-Fermi Kondo model and Kondo lattice model,
only spin appears in the (strongly correlated) local degrees of freedom.
At the corresponding
 Kondo destruction QCP, we find that the charge response not only is singular but also satisfies 
$\omega/T$ scaling. The development of a singular charge response in such a model is surprising,
because the only microscopic charge degrees
of freedom in the Hamiltonian are associated with the non-interacting conduction electrons;
only spins are involved in any of the interaction terms.
It demonstrates the power of quantum (Kondo) entanglement in strongly correlated metallic settings.
More generally, our results capture the aspects of quantum criticality that are unique to
strongly correlated metals, namely the quantum entwining of the charge and spin degrees of freedom.

Our work provides the theoretical basis for the 
understanding of the surprising experimental observation in
Ref.\,\cite{prochaska2018singular},
where
a singular charge response with $\omega/T$ scaling 
is found at 
an antiferromagnetic QCP.
Finally, because Kondo destruction represents a partial-Mott transition,
our work suggests that probing 
the singularities of both charge and spin responses represents a fruitful means of elucidating 
strange metals near an electronic localization, such as the cuprate high temperature superconductors 
and organic charge transfer salts.

We thank K. Ingersent, S. Paschen, and F. Zamani for useful discussions.
Work at Rice was in part supported
by the NSF 
(DMR-1920740)
 and the Robert A. Welch Foundation (C-1411).
S.\ K.\ acknowledges partial support by the National Key R\&D Program of the MOST of China, 
grant No.\ 2016YFA0300200
and the National Science Foundation of China, grant No.\ 11774307.
Computing resources were supported in part by the Data Analysis and Visualization 
Cyberinfrastructure funded by NSF under grant OCI-0959097 and an IBM Shared University 
Research (SUR) Award at Rice University, and by the Extreme Science and Engineering
Discovery Environment (XSEDE) by NSF under Grants No. DMR170109.
Q.S.\ acknowledges 
the hospitality and support by a Ulam Scholarship 
from the Center for Nonlinear Studies at Los Alamos National Laboratory,
and the hospitality of the Aspen
Center for Physics (NSF, PHY-1607611).

\bibliography{SingularCharge}

\end{document}